\documentclass[iop,apjl]{emulateapj}

\begin{document}

\newcommand{\kms}{\ensuremath{\mathrm{km}\,\mathrm{s}^{-1}}}
\newcommand{\MLsun}{\ensuremath{\mathrm{M}_{\sun}/\mathrm{L}_{\sun}}}
\newcommand{\Aunits}{\ensuremath{\mathrm{M}_{\sun}\,\mathrm{km}^{-4}\,\mathrm{s}^{4}}}
\newcommand{\etal}{et al.}
\newcommand{\LCDM}{$\Lambda$CDM}
\newcommand{\ML}{\ensuremath{\Upsilon_*}}


\title{The Baryonic Tully-Fisher Relation of Gas Rich Galaxies
as a Test of \LCDM\ and MOND}

\author{Stacy S. McGaugh}
\affil{Department of Astronomy, University of Maryland}
\affil{College Park, MD 20742-2421, USA}

\begin{abstract}
The Baryonic Tully-Fisher Relation (BTFR) is an empirical relation between baryonic mass and rotation velocity in disk galaxies.  
It provides tests of galaxy formation models in \LCDM\ and of alternative theories like MOND.  Observations of gas rich 
galaxies provide a measure of the slope and normalization of the BTFR that is more accurate (if less precise) than that
provided by star dominated spirals, as their masses are insensitive to the details of stellar population modeling.
Recent independent data for such galaxies are consistent with $M_b = AV_f^4$ with $A = 47\pm6\;\Aunits$. 
This is equivalent to MOND with $a_0 = 1.3 \pm 0.3\;\textrm{\AA}\,\textrm{s}^{-2}$.
The scatter in the data is consistent with being due entirely to observational uncertainties.
It is unclear why the physics of galaxy formation in \LCDM\ happens to pick out the relation predicted by MOND.
We introduce a feedback efficacy parameter $\mathcal{E}$ to relate halo properties to those of the galaxies they
host.  $\mathcal{E}$ correlates with star formation rate and gas fraction in the sense that galaxies that have
experienced the least star formation have been most impacted by feedback.
\end{abstract}

\keywords{galaxies: kinematics and dynamics --- galaxies: dwarf --- galaxies: irregular --- galaxies: spiral}

\section{Introduction}

The Tully-Fisher relation \citep{TForig} is one of the strongest empirical correlations in extragalactic astronomy.
In addition to being a useful distance indicator \citep[e.g.,][]{sakai}, the physical basis of this relation is of
fundamental importance.  In the context of \LCDM, the Tully-Fisher relation places tight constraints on
galaxy formation models 
\citep[e.g.,][]{EL96,MdB98a,MMW98,CR,SN,CR,vdB00, BullockTF,MoMao,MayerMoore,gnedinTF,fabioTF,TGKPR,DRTP,TMZBTDS}.
In a broader context, it provides a useful test of theories that seek to modify gravity in order to obviate the
need for dark matter \citep[e.g.,][]{milgrom83,FLAG,weyl,MoG}.  
These should make specific and testable predictions for the Tully-Fisher relation.

The Tully-Fisher relation was originally posed as an empirical relation between optical luminosity and the 
width of the 21 cm line.  These are presumably proxies for more physical quantities.  
The 21 cm line-width is a proxy for rotation velocity \citep[e.g.,][]{verhTF,YSTF} which
is widely presumed to be set by the dark matter halo.  Luminosity is a proxy for stellar mass, which is
presumably related to the total mass of the system.  Thus a relation like Tully-Fisher seems natural,
with the mass of the parent halo specifying both luminosity and rotation velocity.

Luminosity is not a perfect proxy for mass.  The stellar mass-to-light ratio can, and presumably does, vary
with galaxy type.  This leads to Tully-Fisher relations with different slopes and scatter depending
on the band-pass used to measure the luminosity \citep[e.g.,][]{TFdust,verhTF,Cscale,noordTF,mastersTF}. 
Moreover, low luminosity galaxies tend
to fall below the extrapolation of the fit to bright galaxies \citep{PSnonlinear,MvDG98}.  Yet all these 
observed relations presumably stem from a single underlying physical relation.

\citet{FreemanLSB} and \citet{btforig} suggested that baryonic mass is a more fundamental quantity than 
luminosity.  For any given stellar population, different mass-to-light ratios will be produced in different
bands by the same mass of stars.  More importantly, physics should not distinguish between mass in stars
and mass in other forms.  Adding the observed gas mass to the stellar mass results in a Baryonic Tully-Fisher
Relation (BTFR) that is linear (in log space) over many decades in mass 
\citep{verhTF,BdJ,gurov04,M05,pfennBTF,begum,stark,trach,gurov}.
The BTFR appears to be the fundamental physical relation underpinning the empirical Tully-Fisher relation.
As such, the true form of the BTFR is of obvious importance.

In this paper, I calibrate the BTFR with gas rich galaxies. 
The masses of gas rich galaxies can be more accurately measured than those of the star dominated galaxies
that traditionally define the Tully-Fisher relation because they are less affected by the systematic
uncertainty in the stellar mass-to-light ratio \citep{btforig,stark,newPRL}.
This results in a purely empirical calibration with a minimum of assumptions.

In \S 2, I discuss the data required construct the BTFR, assemble samples of gas rich galaxies, explore the
statistics of the data, and calibrate the relation.  In \S 3 I examine the implications of the relation for \LCDM\ and MOND.
I define a feedback efficacy parameter $\mathcal{E}$ to quantify the mapping between observations and theoretical
dark halo parameters.  I estimate the value of the MOND acceleration constant $a_0$ from the empirical BTFR.
A summary of the conclusions are given in \S 4.

\section{The Baryonic Tully-Fisher Relation}

In order to construct the BTFR, it is necessary to estimate the baryonic mass of a galaxy on the one hand,
and its rotation velocity on the other.  The baryonic mass is the sum of all observed components, stars and gas:
\begin{equation}
M_b = M_* + M_g.
\label{mbary}
\end{equation}
Combining this with measurements of the rotation velocity, one obtains the BTFR:
\begin{equation}
M_b = A V^x.
\end{equation}
Remarkably, this simple formula suffices to describe the data for rotating galaxies.
There is no indication of any need for a third parameter such as 
radius or surface brightness \citep{zwaanTF,sprayTF,CR,myPRL}:  
the Tully-Fisher relation is the optimal projection of the fundamental plane of disk galaxies.

\subsection{Stellar Mass}

The stellar mass of a galaxy is determined from its measured luminosity and an estimate of
its stellar-mass-light ratio \ML:  
\begin{equation}
M_* = \ML L.
\label{mstar}
\end{equation}
The mass-to-light ratio is commonly obtained from stellar population models 
\citep[e.g.,][]{Bell03,portinari}.  These predict correlations between \ML\ and color
so that the observed color is used to estimate \ML.    

The use of a stellar mass-to-light ratio estimated from stellar population models causes 
a significant systematic uncertainty in the stellar mass \citep{conroy}, which is particularly sensitive to the IMF.
Depending on the choice of model, one can derive BTFRs with slopes anywhere in the range 
of $x = 3$ to 4 \citep{M05}.  This is a basic limitation that is hard to avoid in samples of spiral galaxies
where most of the baryonic mass is in the form of stars.  This can be overcome by use of samples
of galaxies whose baryonic mass is dominated by gas rather than stars \citep{stark}.
In such gas rich galaxies, the uncertainty in stellar mass is reduced to a minor contributor to 
the total error budget.  Nevertheless, I am careful to employ a self-consistent population model
to all galaxies employed in the analysis (Table~\ref{datatable}).

\subsection{Gas Mass}
\label{gasmasssect}

Galaxies with $M_g > M_*$ can be found among very late type (Sd and Sm) spirals and Irregulars 
with rotation velocities $V_f < 90\;\kms$ \citep[e.g.,][]{eder,FTHINGS,FIGGS,swaterswhisp}.
With enough such galaxies, it is possible to obtain an absolute calibration of the BTFR
that is effectively independent of the stellar mass estimator \citep{stark}. 
Here I describe gas mass estimates for galaxies used in the analysis.

\subsubsection{Atomic Gas}

Neutral atomic (HI) gas usually dominates the gas component of disk galaxies.  The mass of HI follows directly
from the physics of the spin-flip transition of hydrogen, the cosmic hydrogen fraction, and the distance
to each galaxy.  There is no uncertain mass-to-light ratio as with stars.  

The gas mass is estimated by
\begin{equation}
M_g = \eta M_{HI},
\label{mgas}
\end{equation}
where $M_{HI}$ is the observed atomic gas mass.  
The factor $\eta$ accounts for the cosmic hydrogen fraction and forms of gas other than HI.
Since the majority of objects of interest here are low luminosity late type galaxies, a 
primordial hydrogen fraction is assumed ($X^{-1} = 1.33$) rather than the commonly
assumed solar value ($X^{-1} = 1.4$).  This difference is less than the typical uncertainty in $M_{HI}$, but does
make the numbers here differ slightly from those in the source data.

\subsubsection{Molecular Gas}

After atomic hydrogen, molecular gas is the next significant gas component.
Molecular gas persists in being difficult to detect in late type, low surface brightness galaxies.
Where detected, it appears to contain relatively little of the interstellar gas mass 
\citep{matthewsgao,OHP01,OS04,mathewsCO,dasme,dasmalin2}.
Such direct measurements of molecular gas 
are not available for most galaxies.  To estimate
the molecular gas mass, we might resort to scaling relations for $M_{H_2}/M_{HI}$ such that
\begin{equation}
\eta = \frac{1}{X}{\left(1+\frac{M_{H_2}}{M_{HI}}\right)}.
\label{etaeqn}
\end{equation}
\citet{YK89} showed that $M_{H_2}/M_{HI}$ depends on morphological type, albeit
with large scatter.  From their data, \citet{MdB97} derived the scaling relation
$M_{H_2}/M_{HI} = 3.7 - 0.8 T +0.043 T^2$.  The vast majority of of gas rich galaxies
are very late types ($T \ge 8$), so the typical correction is small ($< 5\%$).

We can use the recent observations of THINGS galaxies \citep{FTHINGS} 
to provide an independent estimate of $M_{H_2}/M_{HI}$.
\citet{leroy} use the observed (low) star formation in dwarf Irregular galaxies to infer the amount
of molecular gas present.  This maps smoothly into the observed molecular content
of brighter galaxies, with all disks following a relation between surface densities such
that $\Sigma_{H_2}/\Sigma_{HI} = \Sigma_*/(81\;\mathrm{M}_{\sun}\,\mathrm{pc}^{-2})$
(their equation 33).  This is a relation between local surface densities, while a relation
between global masses is desired here.  The atomic gas is typically spread over a much
larger area than the stars, so we cannot simply assume that the ratio of surface densities
is the same as the ratio of masses.  However, the size of the stellar and molecular disk 
are usually similar \citep{BIMASONG}.  Assuming that they are the same gives
\begin{equation}
\frac{M_{H_2}}{M_{*}} = \frac{\langle \Sigma_{HI} \rangle}{81\;\mathrm{M}_{\sun}\,\mathrm{pc}^{-2}}.
\label{mh2mhi}
\end{equation}
In this case, the amount of molecular mass depends on the stellar mass rather than the atomic
gas mass and morphological type as in equation~(\ref{etaeqn}).

In order to apply equation (\ref{mh2mhi}), we need an estimate of the average HI surface density.
This is almost always $< 10\;\mathrm{M}_{\sun}\,\mathrm{pc}^{-2}$ in low surface brightness galaxies
\citep{dBMH96}.  In the prototypical case DDO 154, $\Sigma_{HI} \approx 7\;\mathrm{M}_{\sun}\,\mathrm{pc}^{-2}$
at all radii \citep{leroy}.  Taking this value as typical implies that molecular gas is perhaps 9\% of
the stellar mass of these systems.  For galaxies with less than half of their mass in the form
of stars, we again infer that the molecular gas makes up $< 5\%$ of the baryonic mass.

For either of these estimators, the molecular gas content of HI-dominated late type galaxies is small.
So small, in fact, that it is less than the uncertainty in the measurement of the HI mass in most cases.
This would seem to be a good working definition of negligible, so rather than make an uncertain correction
for the molecular gas mass, we neglect it entirely.  In effect, we assume a constant $\eta = 1.33$.

\subsubsection{Other Forms of Gas}

We restrict our baryonic mass budget to the components that are actually detected, as discussed above.
It is of course possible that other reservoirs of baryons exist in some undetected form.
There is, however, no direct evidence for baryons in any form that is comparable in mass to the stars
and atomic gas.

One conceivable reservoir of baryonic mass is hot gas in the halo.  Such a component is expected
in some galaxy formation scenarios.  While some hot gas is certainly present, it appears that its density
is too low to make a substantial contribution here \citep{andersonbregman}.  

Another suggestion is that some of the dark matter could be in the form of undetected cold
molecular gas in galaxy disks \citep{CODM}.  \citet{pfennBTF} showed that a marginal improvement in the scatter
of the BTFR of \citet{btforig} could be obtained by treating $\eta$ as a free parameter, and suggested
that the best fit value ($\eta \approx 3$) implied additional dark baryonic mass in the disk.  
In more accurate data, the scatter is consistent with being caused entirely by observational uncertainties
\citep{verhTF,M05,stark,trach,newPRL}, nullifying this motivation for dark baryons.

In sum, there is no compelling evidence for substantial reservoirs of baryons in disk galaxies
in any form other than those that are directly detected.
I therefore consider only the physically motivated value of $\eta = 1.33$. 
Obviously, a different result would be obtained if we 
allowed $\eta$ to include hypothetical mass components \citep[cf.\ ][]{pfennBTF,begum}.

\subsection{Rotation Velocity}

The characteristic circular velocity of a disk galaxy can be measured with resolved rotation curves,
or approximated by line-widths.  The latter is a flux-weighted integral over the velocity field, and lacks
the clear physical meaning of a rotation curve.  It is, however, readily obtained from single disk 21 cm
observations, whereas obtaining and analyzing interferometric data cubes is considerably more
effort intensive \citep[e.g.,][]{THINGS,trach}.  

When a full rotation curve is available, one can make different choices about where to measure the
circular velocity.  Common choices include the maximum observed  
velocity \citep[$V_{max}$, e.g.,][]{noordTF}, that in
the outermost measured regions where rotation curves tend towards becoming approximately flat
\citep[$V_f$, e.g.,][]{verhTF,M05} or the velocity at some particular optical radius \citep[e.g.,][]{URC,MdB98a,CR}.  
The gas rich galaxies of greatest interest here typically have gradually rising rotation curves that slowly
bend over toward becoming flat without ever having an intermediate peak.  For this common
morphology, all these measures of velocity are very nearly equivalent.  In particular, $V_{max} \approx V_f$,
and the velocity measured at a particular optical radius is also the same unless that radius is chosen
to be very small \citep[see][]{YSTF}.  Indeed, it is not particularly meaningful to define an optical radius for gas
rich galaxies, as the scale length of the baryonic mass can be considerably larger than that of the
stars alone \citep{myPRL,adiabat}.  

For quantitative analysis, I restrict consideration to cases where $V_f$ is explicitly measured,
simply adopting the measured velocity from each source.  For comparison purposes, I also show
line-width ($W_{20}$) data, but do not include these data in the calibration of the BTFR as
line-widths and rotation velocities are comparable but not identical measures.  
In order to make comparisons, I assume $V_f = W_{20}/2$ for gas dominated galaxies with line-width measurements
on the presumption that they would show rising rotation curves like those observed for
comparable galaxies.  For the star dominated galaxies, I assume
$W_{20}/2 = V_p = 1.1 V_f$.  This qualitatively accounts for the fact that bright galaxies have
rotation curves that peak before falling gradually toward flatness \citep{URC}.  The quantitative factor
is chosen to reconcile the BTFR fit to $W_{20}/2$ \citep[$A = 35\;\Aunits$]{btforig} with that fit to $V_f$
\citep[$A = 50\;\Aunits$]{M05}.  This is only done to facilitate comparison between independent samples.

To account for non-circular motions, I accept whatever prescription was used by each source.
In general, these corrections turn out to be rather small \citep[cf.][]{TFcorrections,OhTHINGS,trach,DalcStilp},
as the typical velocity dispersion of low mass galaxies is $\sim 8\;\kms$ \citep{KdN09}.  Since the correction
to the circular velocity occurs in quadrature, it is only
somewhat important only in the slowest rotators.  There exist dynamically cold, rotating disks down to at
least $V_f \approx 30\;\kms$, and perhaps as low as $\sim 10\;\kms$ \citep{tikhonov}.

\subsection{Gas Rich Galaxy Samples}

There is an enormous amount of data in the literature, but only a finite amount of it pertains to gas dominated
galaxies with $M_g > M_*$ that also have all the required information for constructing the BTFR.
I describe here the accumulated data to date.  I also include associated samples of star dominated galaxies.
This is done strictly for comparison as the uncertainties for star dominated galaxies are dominated by the
choice of stellar IMF.  This is the systematic uncertainty that we eliminate by making use of gas dominated galaxies.
For gas rich galaxies, the major sources of uncertainty are the 21 cm flux measurement and the distance to
each galaxy.  The uncertainty in the measurement of $V_f$ is usually much smaller than that in mass (Table~\ref{datatable}),
provided that the inclination is reasonably well constrained.  Cases where this condition might not be satisfied
are discussed and, if suspect, excluded from fitting.

\subsubsection{Stellar Mass Estimates}
\label{stellarmasses}

The data of \citet{begum}, \citet{stark}, and \citet{trach} constitute the sample of gas rich galaxies for which
all the necessary information is available to construct the BTFR.  For these galaxies, stellar mass is, by
definition, a minority of the baryonic mass.  Nevertheless, we take care to estimate 
stellar masses self-consistently.  For specificity, we use their observed $B-V$ colors and
the population model of \citet{portinari} with a Kroupa IMF \citep{kroupa}.  Other plausible choices 
make no difference to the result because the gas mass dominates \citep{stark}.

Several other samples are included for comparison \citep{btforig,M05,gurov}. 
Since these data are not used in the analysis, we make no effort to reconcile them to the same
population model and simply adopt the stellar mass estimated by each source. 
Despite the minor differences in the methods used to estimate stellar mass,
these data are remarkably consistent with one another. 

\subsubsection{Galaxies with Rotation Curve Data}
\label{RCsamples}

\paragraph{\citet{begum}:} Data for remarkably slowly rotating galaxies are presented and analyzed by
\citet{begum,FIGGS}.  The corrections for non-circular motions are substantial in some cases.
However, the most significant difference in the analysis here is that we determine the gas mass as
discussed in \S\ref{gasmasssect}, and we select galaxies for consistency in their inclination determinations.
It is not uncommon for the optical portions of gas rich galaxies to be much smaller than their dominant HI disks
\citep[e.g.,][]{N2915,N3741} and for the outer
HI isophotes to have a different shape than the optical isophotes \citep{OhTHINGS,trach}.   
If the stars suffer an oval distortion
(bar), then an inclination based on optical isophotes will be systematically overestimated \citep{MMdB,dBM98}.  
Indeed, \citet{FIGGS} observe that the optically inferred inclinations often exceed those inferred from HI.
One must be wary that beam smearing might make the HI isophotes rounder than they should be,
potentially causing an underestimate of inclination.  To avoid either effect, 
I demand consistency between optical and HI inclination determinations.
Specifically, I require that $\sin(i_{opt})$ be within 12\% of $\sin(i_{HI})$.  The restriction is placed on $\sin(i)$
because this is what matters to the accuracy with which $V_f$ can be measured.
We are concerned with the velocity at large radii, so the HI inclination is adopted \citep[as in][]{begum}; 
the difference $\Delta i = |i_{opt}-i_{HI}|$ is treated as an additional source of error.  
The particular threshold of 12\% is determined by trial and error, balancing the desire for accuracy with the 
need to retain at least some objects.  With this criterion, 16 objects are retained from Table 1 
of \citet{begum} and 12 are rejected.  Requiring any greater accuracy eliminates nearly the entire
sample.    UGC 5456 is excluded for being marginally star dominated.  Three objects are also in the 
sample of \citet{stark}.

\paragraph{\citet{stark}:} Data for gas rich galaxies are selected from the literature with an emphasis on 
data quality.  \citet{stark} required resolved rotation curves that were observed to flatten sufficiently that the 
logarithmic slope became
$< 0.1$ in their outer parts \citep[see][from which much of this sample originates]{swaterswhisp}.  This
requirement prevents the data from skewing to low $V_f$ due to the inclusion of galaxies whose
rotation curves are still rising rapidly at the last measured point.   

\begin{deluxetable}{lccccr}
\tablewidth{0pt}
\tablecaption{Data for Selected Gas Rich Galaxies}
\tablehead{
\colhead{Galaxy} & \colhead{D} & \colhead{$V_f$} & \colhead{$\log(M_{*})$} & \colhead{$\log(M_g)$} & Ref. \\
& \colhead{Mpc} & \colhead{\kms} & \colhead{M$_{\odot}$}  & \colhead{M$_{\odot}$} & 
}
\startdata
DDO 210	 & \phn0.94       & $\phn17\pm 4\phn$	& $\phn5.88\pm0.15$	& $\phn6.64\pm0.20$	&1,2 \\
Cam B  	 & \phn3.34       & $\phn20\pm12$        & $\phn6.99\pm0.15$	& $\phn7.33\pm0.20$	&1,2 \\
UGC 8215 & \phn4.5\phn & $\phn20\pm 6\phn$	& $\phn6.81\pm0.15$	& $\phn7.45\pm0.20$	&1 \\
DDO 183  & \phn3.24       & $\phn25\pm 3\phn$	& $\phn7.24\pm0.15$	& $\phn7.54\pm0.20$	&1 \\
UGC 8833 & \phn3.2\phn & $\phn27\pm 4\phn$	& $\phn6.94\pm0.15$	& $\phn7.30\pm0.20$	&1 \\
D564-8 	 & \phn6.5\phn & $\phn29\pm 5\phn$	& $\phn6.76\pm0.20$	& $\phn7.32\pm0.13$	&3 \\
DDO 181	 & \phn3.1\phn & $\phn30\pm 6\phn$	& $\phn7.26\pm0.15$	& $\phn7.56\pm0.20$	&1 \\
P51659 	 & \phn3.6\phn & $\phn31\pm 4\phn$	& $\phn6.67\pm0.15$	& $\phn7.85\pm0.20$	&1 \\
KK98 246 & \phn7.83       & $\phn35\pm 6\phn$	& $\phn7.72\pm0.15$	& $\phn7.93\pm0.20$	&1 \\
UGCA 92	 & \phn3.01       & $\phn37\pm 4\phn$	& $\phn7.78\pm0.15$	& $\phn8.32\pm0.20$	&1 \\
D512-2 	 & 14.1\phn       & $\phn37\pm 7\phn$	& $\phn7.58\pm0.20$	& $\phn7.96\pm0.06$	&3 \\
UGCA 444 & \phn0.95       & $\phn38\pm 5\phn$	& $\phn7.34\pm0.15$	& $\phn7.75\pm0.20$	&2 \\
KK98 251	& \phn5.6\phn	& $\phn38\pm 5\phn$	& $\phn 7.34\pm0.15$	& $\phn8.02\pm0.20$	&1 \\
UGC 7242 & \phn5.4\phn & $\phn40\pm 4\phn$	& $\phn7.57\pm0.15$	& $\phn7.78\pm0.20$	&1 \\
UGC 6145 & \phn7.4\phn & $\phn41\pm 4\phn$	& $\phn7.20\pm0.15$	& $\phn7.56\pm0.20$	&1 \\
NGC 3741 & \phn3.0\phn & $\phn44\pm 3\phn$	& $\phn7.24\pm0.15$	& $\phn8.45\pm0.20$	&1,2 \\
D500-3   & 18.5\phn       & $\phn45\pm 6\phn$	& $\phn6.97\pm0.20$	& $\phn7.94\pm0.05$	&3 \\
D631-7   & \phn5.5\phn & $\phn53\pm 5\phn$	& $\phn6.88\pm0.20$	& $\phn8.29\pm0.15$	&3 \\
DDO 168  & \phn4.3\phn & $\phn54\pm 3\phn$	& $\phn8.07\pm0.15$	& $\phn8.74\pm0.20$	&2 \\
KKH 11   & \phn3.0\phn & $\phn56\pm 5\phn$	& $\phn7.28\pm0.15$	& $\phn7.85\pm0.20$	&1 \\
UGC 8550 & \phn5.1\phn & $\phn58\pm 3\phn$	& $\phn8.25\pm0.37$	& $\phn8.46\pm0.39$	&2 \\
D575-2   & 12.2\phn       & $\phn59\pm 7\phn$	& $\phn7.63\pm0.20$	& $\phn8.62\pm0.07$	&3 \\
UGC 4115 & \phn7.5\phn & $\phn59\pm 6\phn$	& $\phn7.77\pm0.15$	& $\phn8.58\pm0.20$	&1 \\
UGC 3851 & \phn3.2\phn & $\phn60\pm 5\phn$	& $\phn8.45\pm0.15$	& $\phn9.09\pm0.20$	&2 \\
UGC 9211 & 12.6\phn       & $\phn64\pm 5\phn$	& $\phn8.12\pm0.39$	& $\phn9.21\pm0.41$	&2 \\
NGC 3109 & \phn1.3\phn & $\phn66\pm 3\phn$	& $\phn7.41\pm0.15$	& $\phn8.79\pm0.20$	&2 \\
UGC 8055 & 17.4\phn       & $\phn66\pm 7\phn$	& $\phn8.09\pm0.15$	& $\phn9.02\pm0.20$	&1 \\
D500-2   & 17.9\phn       & $\phn68\pm 7\phn$	& $\phn7.41\pm0.20$	& $\phn9.06\pm0.05$	&3 \\
IC 2574  & \phn4.0\phn & $\phn68\pm 5\phn$	& $\phn8.94\pm0.15$	& $\phn9.20\pm0.20$	&2 \\
UGC 6818 & 18.6\phn       & $\phn72\pm 6\phn$	& $\phn9.22\pm0.16$	& $\phn9.28\pm0.20$	&2 \\
UGC 4499 & 13.0\phn       & $\phn74\pm 3\phn$	& $\phn8.75\pm0.27$	& $\phn9.32\pm0.29$	&2 \\
NGC 1560 & \phn3.45       & $\phn77\pm 3\phn$	& $\phn8.70\pm0.15$	& $\phn9.23\pm0.20$	&2 \\
UGC 8490 & \phn4.65       & $\phn78\pm 3\phn$	& $\phn8.36\pm0.15$	& $\phn8.96\pm0.20$	&2 \\
UGC 5721 & \phn6.5\phn & $\phn79\pm 3\phn$	& $\phn8.17\pm0.37$	& $\phn9.05\pm0.39$	&2 \\
F565-V2  & 48.\phn\phn & $\phn83\pm 8\phn$	& $\phn8.30\pm0.21$	& $\phn9.04\pm0.24$	&2 \\
F571-V1  & 79.\phn\phn & $\phn83\pm 5\phn$	& $\phn9.00\pm0.19$	& $\phn9.33\pm0.22$	&2 \\
IC 2233  & 10.4\phn       & $\phn84\pm 5\phn$	& $\phn8.96\pm0.15$	& $\phn9.32\pm0.20$	&2 \\
NGC 2915 & \phn3.78       & $\phn84\pm10$	        & $\phn7.99\pm0.15$	& $\phn8.78\pm0.20$	&2 \\
NGC 5585 & \phn5.7\phn & $\phn90\pm 3\phn$	& $\phn8.98\pm0.38$	& $\phn9.27\pm0.40$	&2 \\
UGC 3711 & \phn7.9\phn & $\phn95\pm 3\phn$	& $\phn8.92\pm0.15$	& $\phn9.01\pm0.17$	&2 \\
UGC 6983 & 18.6\phn       & $108\pm3\phn$	        & $\phn9.53\pm0.16$	& $\phn9.74\pm0.20$	&2 \\
F563-V2  & 61.\phn\phn & $111\pm5\phn$	        & $\phn9.41\pm0.17$	& $\phn9.63\pm0.21$	&2 \\
F568-1   & 85.\phn\phn & $118\pm4\phn$	        & $\phn9.50\pm0.18$	& $\phn9.87\pm0.22$	&2 \\
F568-3   & 77.\phn\phn & $120\pm6\phn$	        & $\phn9.62\pm0.18$	& $\phn9.71\pm0.22$	&2 \\
F568-V1  & 80.\phn\phn & $124\pm5\phn$	        & $\phn9.38\pm0.18$	& $\phn9.65\pm0.22$	&2 \\
NGC 2403 & \phn3.18       & $134\pm3\phn$	        & $\phn9.61\pm0.15$	& $\phn9.77\pm0.20$	&2 \\
NGC 3198 & 14.5\phn       & $149\pm3\phn$	        & $10.12\pm0.15$	& $10.29\pm0.20$	& 2 \\
\enddata
\tablerefs{1. \citet{begum}; 2. \citet{stark}; 3. \citet{trach}.}
\tablecomments{An electronic version of this table is available at 
http://www.astro.umd.edu/\~{}ssm/data/gasrichdatatable.txt}
\label{datatable}
\end{deluxetable}

\paragraph{\citet{trach}:} New WSRT observations of 11 very gas rich galaxies selected from the sample
of \citet{gasrich} and \citet{eder} are used to measure both $W_{20}$ and $V_f$.  Of the 11 galaxies
observed, 6 show clear evidence of rotation.  Interestingly, these all have HI profiles that are
double-horned or at least show some structure.
Galaxies for which rotation is not positively detected tend to have single-horned
profiles.  \citet{trach} discuss two methods for measuring the outer circular velocity that are largely consistent.  
I adopt for $V_f$ their values measured from the position-velocity diagram ($V_{pv}$ from their Table~3).
For D575-2, I estimate $V_f = 59\pm7\;\mathrm{km}\,\mathrm{s}^{-1}$ 
based on the outermost detected 21 cm emission in their data.  

\paragraph{\citet{M05}:} The data for star dominated galaxies with $V_f$ measured from extended
HI rotation curves are shown for comparison.  They are not used in the analysis here. 
For illustration, stellar masses adopt the ${\cal Q} = 1$ MOND mass-to-light ratios.
The BTFR fit to these data [$M_b = (50\;\Aunits) V_f^4$] can be tested by extrapolation to the much lower circular velocities
of the new gas rich galaxies.

\begin{figure*}
\epsscale{1.0}
\plotone{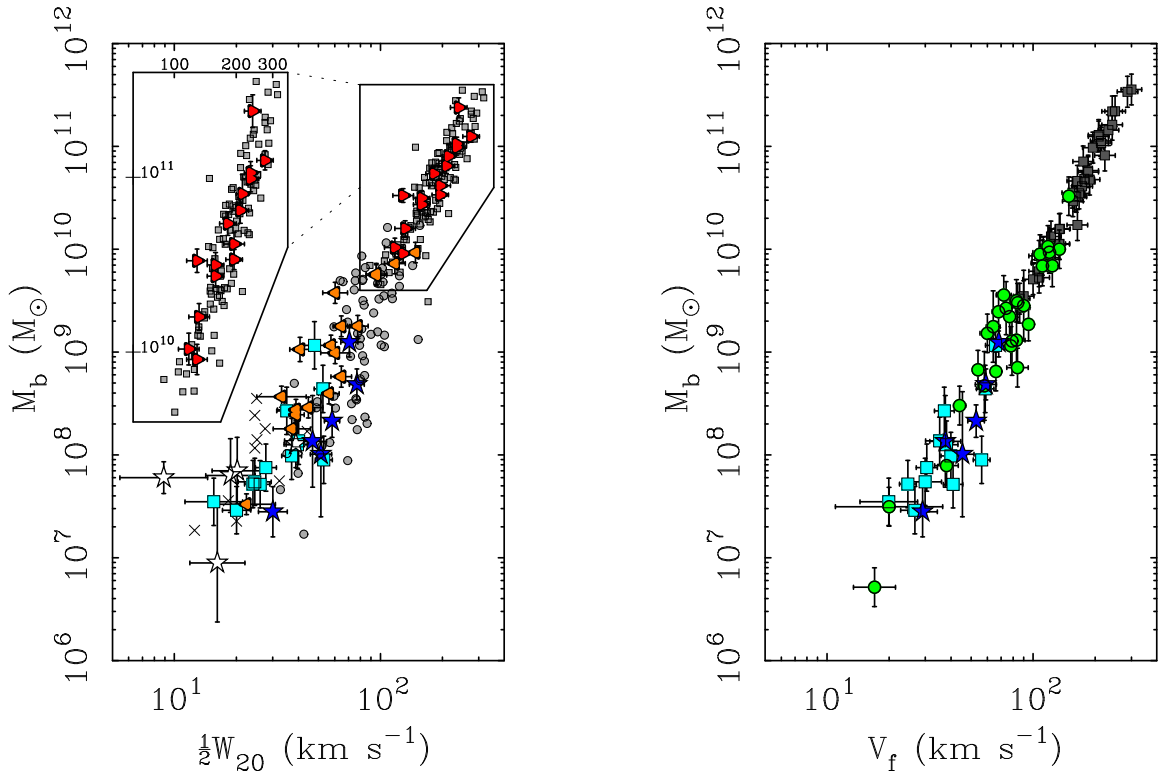}
\caption{Baryonic mass as a function of line-width (left) and circular velocity as measured
from resolved rotation curves (right).  
Left panel:  data from \citet{btforig} are shown as light gray points with uncertainties suppressed for clarity.
Gas dominated galaxies (with $M_* < M_g$) are shown as small circles 
and star dominated galaxies ($M_* > M_g$) are shown as small squares.  
More recent data are shown with larger symbols.
The data of \citet{trach} are shown as dark blue stars in cases where 
rotation is detected and as open stars when it is not.  The data of \citet{begum} are shown as
light blue squares for cases with consistent optical and HI inclinations and as crosses in cases where
these differ substantially.  The HI data of \citet{gurov} are shown as orange left-pointing triangles.  
The data of the star dominated galaxies of \citet{sakai} are shown as red right-pointing triangles
with stellar mass from the $H$-band data of \citet{gurov}.  The two samples of star dominated galaxies 
are compared in the inset.  This also serves to illustrate the range covered by the bright spirals that 
traditionally define the Tully-Fisher relation --- a range that is vastly expanded by the inclusion of gas
rich dwarfs.  In the right panel, data for galaxies with $V_f$
measured from resolved rotation curves include the rotating cases of \citet{trach}, 
the data of \citet{begum} with consistent inclinations, 
the gas dominated galaxies compiled by \citet[green circles]{stark}, and the star dominated
galaxies compiled by \citet[dark gray squares]{M05}.  
\label{btffancyseb}}
\end{figure*}

\placefigure{btffancyseb}

The three data sets of \citet{begum}, \citet{stark},and \citet{trach} are the most interesting for our purposes here.
They compose a sample of gas dominated galaxies with resolved rotation curve measurements.
These data are listed in Table~\ref{datatable} and are used in the analysis.  The data for these
galaxies is less precise than those for bright star dominated galaxies.  However, they are more
accurate, as their masses are not sensitive to systematic uncertainties in the IMF and stellar population
mass-to-light ratios.

\subsubsection{Galaxies with Line-width Data}

Samples of galaxies with line-width data are used for comparison purposes only.  We shall see that line-width data give a systematically
shallower slope and greater scatter for the BTFR than do rotation curve data.  This appears to be a limitation of the line-width as
a proxy for the rotation velocity.  Aside from this detail, the data present a consistent overall picture.

\paragraph{\citet{btforig}:} Gas rich galaxies were selected from the sample of \citet{gasrich} and \citet{eder} 
to have inclinations $i > 45\degr$.  Stellar masses were estimated from $I$-band data assuming a 
constant mass-to-light ratio ($\ML^I = 1.7\;\MLsun$).  
We retain this original prescription, making only the correction to stellar mass discussed by \citet{gurov04}.  
Data for star dominated galaxies with $H$-band photometry 
were taken from \citet{nuts}.  A constant $H$-band mass-to-light ratio of $\ML^H = 1\;\MLsun$ was assumed.
These data are included here for comparison to more recent data but are not used in the analysis.

\paragraph{\citet{gurov}:} New data are presented for galaxies with $M_g > M_*$.
Detailed stellar population modeling has been performed to determine the stellar mass 
\citep[Table 11 of][]{gurov}.  The same modeling procedure has also been applied to the star
dominated sample of \citet{sakai}.  We adopt the stellar mass determinations based on $H$-band 
photometry\footnote{Whether optical or $H$-band data are used to estimate the stellar mass for the
bright galaxies of \citet{sakai} makes a difference to the slope of the BTFR: $x = 3$ in the optical and $x = 4$
in the $H$-band.  This exemplifies the systematic uncertainty that gas dominated galaxies save us from.}
for these galaxies \citep[Table 4 of][]{gurov} to facilitate direct comparison with the data of \citet{nuts}
employed by \citep{btforig}.  All of the bright galaxies of \citet{sakai} have nicely double horned
line profiles with well measured line-widths.  This is not the case for the new gas rich galaxies presented by \citet{gurov},
for which the HI profiles  are all narrow and single-horned (Gurovich, private communication).  Such profiles can be
poor indicators of $V_f$ \citep{trach}.

\subsection{Comparison of the Data}
\label{datacomp}

Fig.~\ref{btffancyseb} shows the data.  Over all, the agreement between the various data is good.
This is true even for the star dominated galaxies, despite the differences in the population modeling.
The spiral galaxies of \citet{sakai} fall in the midst of the data of \citet{nuts} (inset of Fig.~\ref{btffancyseb})
with indistinguishable slope and normalization.  The chief difference is the lower scatter in the data 
of \citet{sakai}.  This presumably results from the more accurate distances to the individual galaxies
based on HST observations \citep{HSTKP}.  This is expected:  as the data improve, the scatter is reduced.
Indeed, essentially all of the scatter can be accounted for by observational uncertainty and the expected
variation in stellar mass-to-light ratios \citep{verhTF,M05,stark}.

\begin{figure*}
\epsscale{1.0}
\plotone{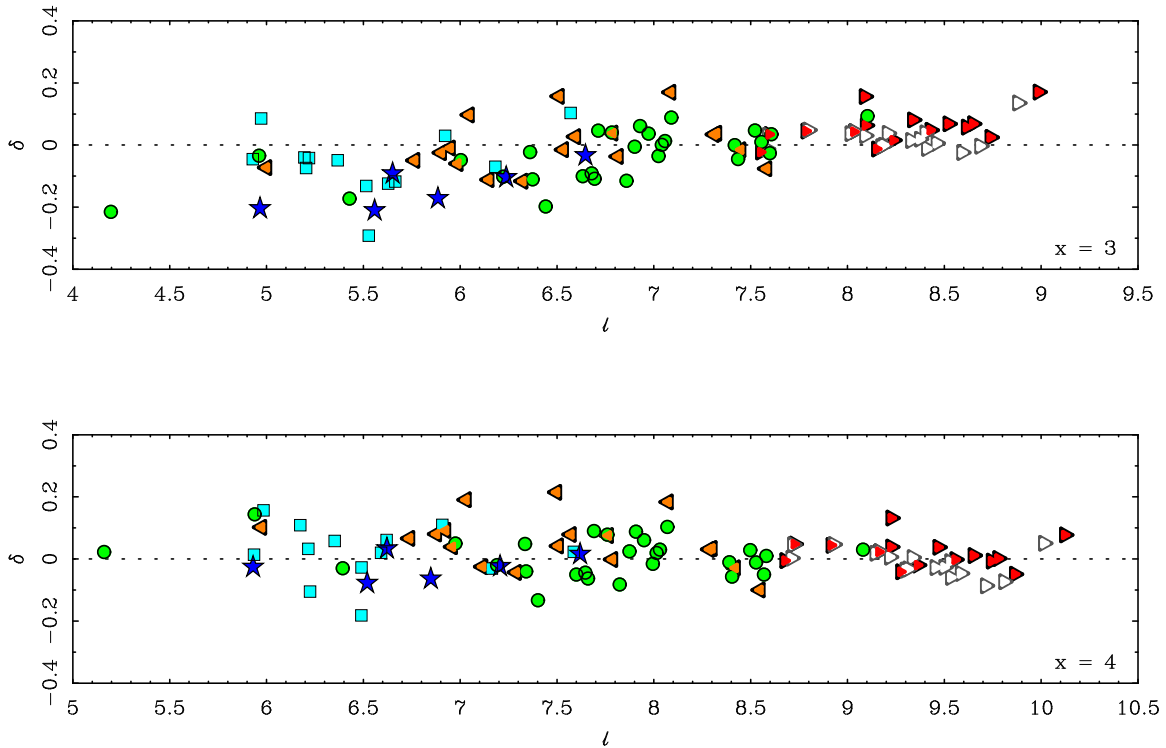}
\caption{The data in a coordinate system rotated (equations \ref{eqn:rotatea} and \ref{eqn:rotateb}) so that
the abscissa is the position of a galaxy along the BTFR and the ordinate is its perpendicular distance
from the BTFR.  Symbols as per Fig.~\ref{btffancyseb} with the exception that the star dominated galaxies
of \citet{sakai} are shown twice:  once with the $H$-band stellar masses of \citet{gurov} as in Fig.~\ref{btffancyseb}
and again for the hybrid $V$ and $H$ band stellar masses of \citet[open right-pointing triangles]{gurov}.
This illustrates the systematic uncertainty that plagues star dominated galaxies but from which gas dominated
galaxies are largely immune.  The top panel shows the BTFR with slope 3 fit to the data of \citep{gurov}.
The bottom panel shows the BTFR fit of \citet{M05} with slope 4.  
The data shown in the bottom panel are independent of those to which the fit was made.  
\label{rotate}}
\end{figure*}

\placefigure{rotate}

The scatter is larger for lower luminosity galaxies.  This is also expected: it is more challenging
observationally to obtain precise data for faint dwarfs than it is for bright spirals.  Nevertheless, the
data for the gas rich dwarfs appear to be largely consistent.  Moreover, the scatter is reduced as
we progress from line-widths to resolved measures of the rotation velocity (left vs.\ right panels in
Fig.~\ref{btffancyseb}).  The improved accuracy with which the rotation velocity is measured is
reflected in reduced scatter.

There are some outliers among the gas rich dwarfs, especially when $W_{20}$ is used as the
velocity estimator.  The two most dramatic outliers from the sample of \cite{begum}
 are UGC 3755 and KK98-65, which fall
far to the left of the main correlation.  These are rejected by the requirement of consistent optical and HI inclinations
as are a number of galaxies that do lie along the BTFR in Fig.~\ref{btffancyseb} when the HI inclination is used.
The outliers from \citet{trach} are those objects for which rotation is not clearly detected.
(D575-5 is the most discrepant case).  Three objects from the sample of \citet{gurov} stand out 
from the rest of the data:  ESO 085-G 088, AM 0433-654, and HIPASS J1424-16b.  ESO 085-G 088
has a large velocity error, so its status as an outlier is only marginal.  The other two objects clearly 
deviate from the rest of the data, including the bulk of the data of \citet{gurov}.  
Since we do not have resolved measures of rotation for these objects, it is possible that they, like the
single-peaked HI profiles of \citet{trach}, would not show clear evidence of rotation if so observed.
Another possibility is that the optical inclinations used by \citet{gurov} are systematically in error
as discussed in \S\ref{RCsamples}.  Lacking resolved HI observations, we can not make the same
consistency check that is possible with the data of \citet{begum}.  
The inclination of AM 0433-654 is $43\degr$ and that of HIPASS J1424-16b is $47\degr$.
A shift of $\sim 0.15$ dex in $W_{20}$ would reconcile these galaxies with the other data.  
If inclination is responsible, the corresponding inclinations would be $ 29\degr$ for AM 0433-654 
and $31\degr$ for HIPASS J1424-16b.  From experience with
such objects \citep[e.g.,][]{dBM98,trach}, this is well within the realm of possibility.  Indeed, it would be surprising
if something like this did not happen with only optical inclinations to work with.  

Over all, the data paint a consistent picture (Fig.~\ref{btffancyseb})
despite the inevitable concern over the accuracy of a few individual measurements.  
The BTFR is continuous and linear in log space over five decades in mass.
There is a hint of curvature when $W_{20}$ is used as the rotation velocity measure,
but it is not statistically significant.  There is not even a hint of curvature when resolved
rotation curves are available to measure $V_f$.

We expect the accuracy of the data to improve as we move from integral measures like the line-width to
physically meaningful rotation velocities measured from resolved HI data cubes.  This expectation is
realized in the reduced scatter seen in the right panel of Fig.~\ref{btffancyseb}.  
We can quantify this with ordinary least-square fits to the data.  

If we apply the maximum likelihood method of \citet{weiner} to fit the new line-width data 
\citep[i.e., all the data in the left panel of Fig.~\ref{btffancyseb} except those of][]{btforig},
we obtain a slope $x = 3.41 \pm 0.08$ and an intercept $\log A = 2.91 \pm 0.17$.
If we perform the same exercise with the rotation curve data in the right panel of Fig.~\ref{btffancyseb},
we obtain $x = 3.94 \pm 0.11$ and $\log A = 1.81 \pm 0.22$.  These formal fits have rather different slopes
even though the chief difference is in the scatter.    The method of \citet{weiner} accounts for uncertainties
in both velocity and mass, and can optionally estimate the amount of intrinsic scatter in a sample.
Accounting for the stated observational errors
\citep[e.g.,][]{weiner} implies an intrinsic scatter in the line-width data of $\sim 0.2$ dex.
That is, there is 0.2 dex of scatter not explained by observational uncertainty.
In contrast, this number is zero for the rotation curve data in the right hand panel:
observational errors suffice to explain all the scatter, leaving no room for
intrinsic scatter.  This difference in scatter implies some source of variance in line-width measures 
that is perhaps not accounted for in the stated uncertainties, and not likely to be meaningful to the physics
we are interested in here.

At least part of the reason for the difference in fitted slopes is physical:  $W_{20}$ is not a perfect proxy for $V_f$.  
Bright galaxies typically have peaked rotation curves with $V_{max} > V_f$.  Line-width measurements
do not resolve where emission originates, and are thus sensitive to the maximum velocity rather than
that at larger radius.  This immediately skews the slope to shallower values.  A similar if less pronounced
effect can be seen in optical rotation curves that are resolved but not particularly extended \citep{ghasp}.

For the physical interpretation of the BTFR, we wish to have as precise a probe as possible of the quantity
of interest.  For \LCDM, this would be the virial velocity of the dark matter halo.  In the case of MOND, 
we need a measure that probes the low acceleration regime.  In both cases, $V_f$ provides a better measure
than $W_{20}$.

\subsection{Statistics of the Gas Rich Galaxy Data}
\label{SSSK}
\subsubsection{Slope}

The slope of the BTFR has been addressed by many authors 
\citep[e.g.,][]{btforig,BdJ,gurov04,M05,begum,stark,gurov}, and is typically found to be between
$x = 3$ and 4.  Most of this variation is due to the prescription adopted for estimating the mass-to-light
ratio \citep{M05}, which is by far the strongest systematic effect.  This can be
remedied by selecting gas dominated galaxies as described in \S\ref{stellarmasses}.

\citet{stark} found a slope $x = 3.94 \pm 0.07$ with a bisector fit while \citet{gurov} found
$x = 3.0 \pm 0.2$ from their gas dominated galaxies (their Table 12).  These results appear to be contradictory.
Repeating the analysis of \citet{gurov}, I obtain the same result.  However, simply looking at the data,
it is not obvious that they are in conflict.  Part of the difference is simply the difference in velocity estimators,
as line-widths give shallower slopes than resolved rotation velocities.  Their sample is also small, so
is easily influenced by a few outliers as discussed above (\S \ref{datacomp}).

This difference between
$x \approx 3$ and 4 is precisely the difference between the nominally anticipated slopes in 
\LCDM\ and MOND.  The correct answer is therefore of considerable importance and lends itself
to controversy.  Bear in mind that the data themselves are quite consistent
(Fig.~\ref{btffancyseb}), so any controversy is in the fitting and choice of velocity estimator, not in the data.
The line-width is a convenient proxy for the rotation velocity when resolved rotation curves are unavailable, 
but not a substitute for them when predictive theories are being tested.

\begin{figure*}
\epsscale{1.0}
\plotone{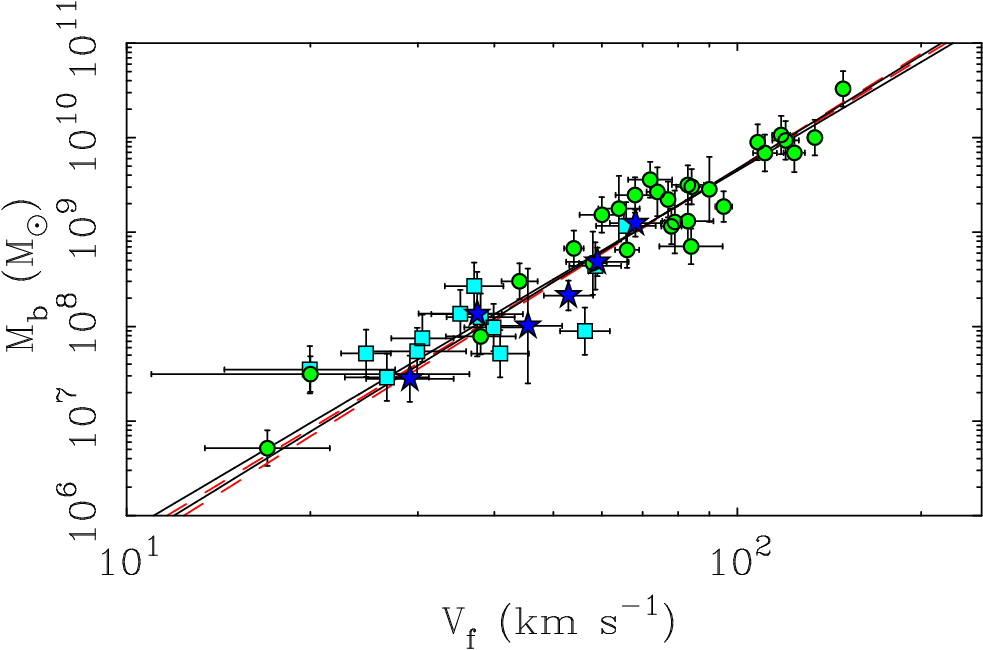}
\caption{Fits to the BTFR of gas dominated galaxies.  Symbols as per Fig.~\ref{btffancyseb}.
The solid lines show the best fit to the whole sample (with slope $x = 3.82$) and the bisector (slope 3.98)
of the forward and reverse fits (dashed lines) limited to the samples of \citet{stark} and \citet{trach} --- 
that is, excluding the less accurate points (squares) from \citet{begum}.  The various fits are consistent within the uncertainties,
and yield the same basic result.
\label{btfgasfit}}
\end{figure*}

\placefigure{btfgasfit}

To illustrate the difference between slopes of 3 and 4, we rotate the BTFR into a new coordinate system
($\ell$,$\delta$) parallel to it such that $\ell$ measures the position of a galaxy along the BTFR and
$\delta$ its deviation perpendicular to it.  That is,
\begin{eqnarray}
\ell = \cos\theta \log V_f + \sin\theta (\log M_b - \log A)
\label{eqn:rotatea} \\
\delta = \cos\theta (\log M_b - \log A) - \sin\theta \log V_f
\label{eqn:rotateb}
\end{eqnarray}
where $\tan\theta$ is the slope of the BTFR.  In Fig.~\ref{rotate} we show the data rotated into the
frames defined by the fits of \citet{gurov} with slope $\tan\theta = 3$ and \citet{M05} with $\tan\theta = 4$.

A slope of 4 does a good job of describing the bulk of the data:
there is no systematic deviation from the dotted line in the lower panel of Fig.~\ref{rotate}.
In contrast, there is an obvious residual from a slope of 3 in the the top panel.
The data show a clear preference for the steeper slope.

Fig.~\ref{rotate} also illustrates the systematic caused by different mass-to-light ratio estimates for star
dominated galaxies.  The galaxies from \citet{sakai} are shown twice, once with masses based on $H$-band data
\citep[Table 4 of][]{gurov} and again with the average of $V$ and $H$ band stellar masses 
\citep[Table 5 of][]{gurov}.  The hybrid $V$-$H$ masses are nicely consistent with a slope of 3,
while $H$ alone prefers a slope of 4.  While one must choose which band to trust for bright galaxies, 
this choice is rendered irrelevant for gas dominated galaxies.  We therefore concentrate the rest of our 
analysis on the gas rich galaxy sample described in \S\S\ref{RCsamples} and \ref{stellarmasses}.

Taking the 47 gas rich galaxies with reasonably trustworthy data (Table~\ref{datatable}), the best fit slope 
is $x = 3.82 \pm 0.22$ with intercept $\log A = 2.01 \pm 0.41$.  Various methods (forward, reverse,
and maximum likelihood) all give the same result.  Of these data, the velocity measurements of \citet{begum} are
the most challenging: only three of their objects meet the quality criteria of \citet{stark}.  Excluding the data of \citet{begum}
but retaining those of \citet{stark} and \citet{trach} gives $x = 4.05 \pm 0.29$ from a forward fit, $3.92 \pm 0.29$ from a
reverse fit, and a bisector slope \citep{isobe} of $3.98 \pm 0.06$.  The corresponding intercepts are $\log A = 1.57$, 1.82, and 1.71,
respectively, with uncertainty $\pm 0.24$.  The slight difference that occurs when also fitting the data of \citet{begum}
may result from a slight skew to low velocities, and testifies to the difficulty in obtaining quality data for very slow
rotators \citep[see also][]{trach}.  The skew effect becomes more pronounced if we
relax the selection criterion that requires consistent optical and HI inclination determinations.  Irrespective of these details, 
these fits are all consistent with one another (Fig.~\ref{btfgasfit}), and none yield a slope that is meaningfully different from 4.

\subsubsection{Normalization}

An interesting question we can pose to the gas rich galaxy data is what the normalization of the BTFR is with slope fixed to 4.  
The gas rich galaxies provide an absolute calibration of the BTFR that is very nearly independent of the estimator
we use for the stellar mass-to-light ratio.  By fixing the slope, we can sharpen our estimate of 
the intercept, and also examine higher order moments of the distribution.  In addition to the scatter perpendicular
to the BTFR, $\sigma_{\delta}$, we can also examine the skew $\alpha_3$ and kurtosis $\hat \alpha_4$.

Table \ref{fittab} shows these statistics for several combinations of the rotation curve samples.  We start with
the sample of \citet{stark} as their sample was constructed for this exercise.  Adding the galaxies of \cite{trach}
decreases the intercept slightly.  Adding the data of \citet{begum} 
raises again, back near to the value found by \citet[$\log A = 1.70$]{M05}.  

Table~\ref{fittab} gives quantitative estimates for the intercept.  For each combination of samples both a weighted
and unweighted estimate is given.  The weighted estimate comes from minimizing $\chi_{\nu}^2$ by adjusting $\log A$
with fixed slope:  we minimize the orthogonal deviations.  The unweighted estimate is made by finding the centroid of 
the distribution in the orthogonal deviations $\delta$ of the rotated data
and requiring the centroid to lie at $\delta = 0$.  The latter method gives no heed to the accuracy of the
data, but is less sensitive to the influence of outliers whose formal uncertainty may understate systematic effects.
These methods give indistinguishable results.

\begin{deluxetable*}{lrccccccc}
\tablewidth{0pt}
\tablecaption{Statistics of Gas Rich Galaxy Rotation Curve Samples}
\tablehead{
\colhead{Sample} & \colhead{N} & \colhead{$\chi_{\nu}^2$} & \colhead{$\log A$} & \colhead{$\log A$} &
\colhead{$\alpha_3$} & \colhead{$\hat \alpha_4$} & \colhead{$a_0$} & \colhead{$a_0$} \\
& & & \colhead{weighted} & \colhead{unweighted} & skew & kurtosis &
 \colhead{weighted} & \colhead{unweighted}  
}
\startdata
S       &  28   & 1.04  & 1.71 &  1.65  & $-$0.07 & $-$0.14 &  1.18 & 1.36 \\
ST     &  34   & 0.92  & 1.67 &  1.64  & $-$0.01 & \phs0.01 & 1.30 & 1.39 \\
STB   &  47  & 0.99  & 1.69 &  1.71  & $-$0.13 & $-$0.10 &  1.24 & ~~1.18 
\enddata
\tablecomments{Sapmles: S: \citet{stark}; T: \citet{trach}; B: \citet{begum}.
The units of $A$ are $\Aunits$ and those of $a_0$ are $\mathrm{\AA}\;\mathrm{s}^{-2}$.
The skew and kurtosis are the statistics of the orthogonal deviations $\delta$ (equation \ref{eqn:rotateb}).}
\label{fittab}
\end{deluxetable*}

All combinations of the various samples give consistent estimates for the intercept.  
To be specific, we adopt the weighted estimate from the combination of the Stark-Trachternach sample:
\begin{equation}
A = 47 \pm 6\;\Aunits.
\label{bestfitA}
\end{equation}
This is the midpoint of the various determinations and has the smallest $\chi^2$.
This value is of course very similar to the unweighted value ($45\;\Aunits$) obtained from the \citet{stark} 
sample alone \citep{M10,MWolf}.  It is also similar to the value ($50\;\Aunits$) found by \citet{M05} from star-dominated
galaxies, but has the virtue of being completely empirical with little sensitivity to how stellar masses are estimated.

It appears that at this juncture, systematic effects and the necessary assumptions are as limiting as data quality.
It is traditional to assume $\eta = 1.4$ which is appropriate for solar metallicity atomic gas, but is slightly larger than the primordial
1.33 taken here for these little evolved, gas rich, low mass galaxies.  The choice of $\eta = 1.4$ is not important to
the star dominated galaxies in \citet{M05}, but we note that adopting it here would cause a change in mass of at most 5\%
(for purely gaseous galaxies) that would drive $A = 47 \rightarrow 49\;\Aunits$.  This is at the level of our best
guess in the mass of molecular mass.  At this level the systematic uncertainty in the IMF that plagues stellar
mass estimates also begins to play a role even in gas dominated galaxies.  Given these considerations, any
value in the range 45 --- $50\;\Aunits$ is plausible, with values much outside this range quickly becoming less plausible.
If we use the BTFR to estimate stellar masses \citep{stark}, then more massive star dominated galaxies are consistent
with a Kroupa IMF.  Grossly different stellar masses would imply a break in the slope of the relation around $\sim 100\;\kms$
where the galaxy population transitions from frequent to infrequent gas domination.

\subsubsection{Scatter}

The scatter in our rotated coordinates is $\sigma_{\delta} = 0.06$ with only the third digit varying from sample to sample.
In terms of mass this is 0.24 dex.  The bulk of this uncertainty (0.20 dex) is provided by the error in the baryonic
mass.  Velocity uncertainties are less important in most samples, and only rival the uncertainty in mass in the lowest
mass systems.  Together, the uncertainties in mass and rotation velocity suffice to account for essentially all of the observed 
scatter.  

\begin{figure*}
\epsscale{1.0}
\plotone{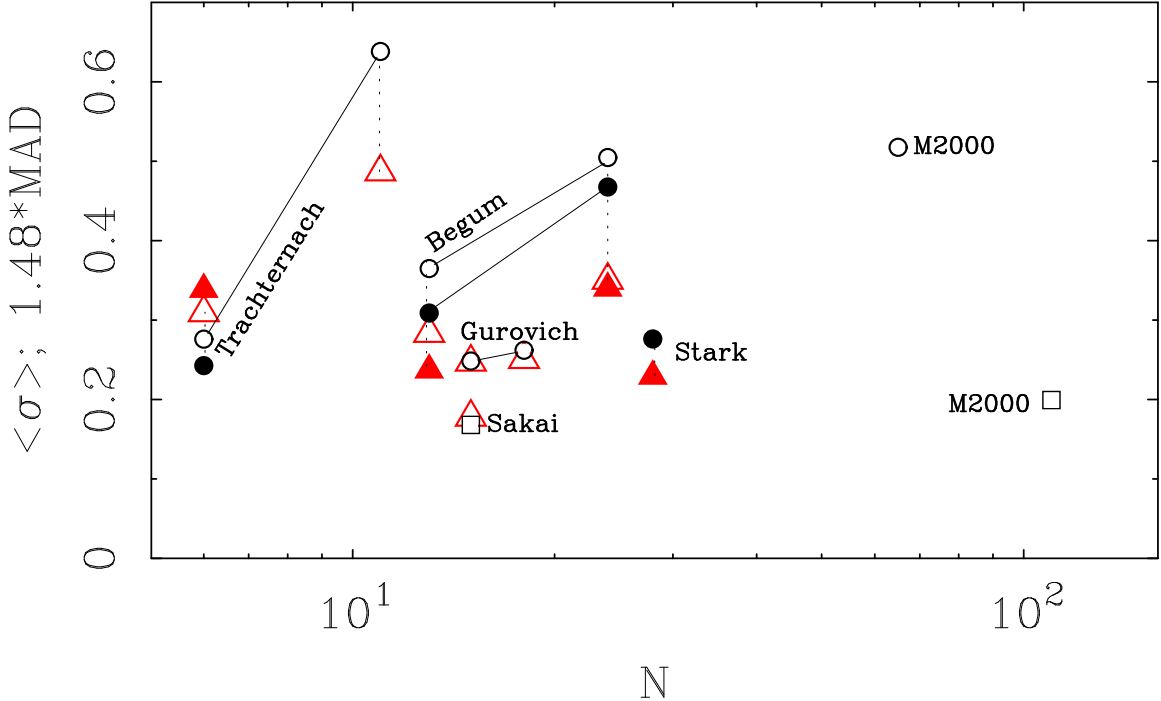}
\caption{The median error (triangles) and median absolute deviations (MAD) of the various samples.
Open symbols refer to line-width measurements; solid symbols represent rotation curve measures.
Lines connect samples that have been trimmed for quality as discussed in the text, showing both trimmed
and untrimmed samples.  Samples are labeled by first author with the MAD of gas dominated galaxies represented
by circles and that of star dominated galaxies by squares.   
The MAD is a robust estimator of the dispersion of a sample; for a normal distribution $\sigma = 1.48$MAD.
The similarity of the observed scatter to the typical uncertainty is an indication that the BTFR has very little intrinsic scatter.
\label{scatter}}
\end{figure*}

A robust estimator of the dispersion of a sample is the median absolute deviation (MAD).  
The dispersion of the various samples about the BTFR as measured by their MAD is shown in Fig.~\ref{scatter}. 
Also shown is the median measurement uncertainty $\langle \sigma \rangle$.  For data that scatter about the
relation simply because of observational uncertainty, $\langle \sigma \rangle \rightarrow$ 1.48MAD as N becomes
large presuming there are no systematic uncertainties that have not been accounted for.  Fig.~\ref{scatter} shows
that indeed $\langle \sigma \rangle \approx$ 1.48MAD in most cases.  This leaves little, if any, room for intrinsic
scatter in the BTFR.  Indeed, the only times when scatter measured by the MAD significantly exceeds that expected
from measurement errors are in the cases where we have identified systematic uncertainties that are not included
in the error estimate:  in the subset of the sample of \citet{trach} where no signature of rotation has been detected,
and in the sample of \citet{begum} where the optical and HI inclination estimates differ greatly.  Had we been unaware
of these sources of systematic error, we might infer some finite intrinsic scatter.  Instead, it appears that including less
accurate data merely increases the scatter as one would expect.

The BTFR of gas rich galaxies appears to have essentially zero intrinsic scatter. 
The same conclusion follows from noting that $\chi_{\nu}^2$ is close to unity (Table~\ref{fittab}).  
A finite intrinsic width to the relation would cause $\chi_{\nu}^2 > 1$ \citep{weiner}.
We can place an upper limit on the scatter by assuming the data are Gaussian (a dubious but conservative
assumption for this purpose) and subtracting the measurement error in quadrature from 1.48MAD.  
This gives a limit on the intrinsic scatter of $\sigma_{int} < 0.15$ dex in mass for the \citet{stark} sample and 
$\sigma_{int} < 0.18$ for the trimmed sample of \citet{begum}.  Essentially zero intrinsic scatter is inferred by
this method for the gas rich data of \citet{gurov} or the star dominated galaxies of \citet{sakai}.  For the data of
\citet{trach}, the expected error exceeds the observed scatter.  This might indicate that the uncertainties
(especially in distance) have been overestimated, but the sample is small so we should not expect the
assumption of a normal distribution to hold.  Indeed, that is generally true here, which mitigates against
there being much intrinsic scatter.

\placefigure{scatter}

It is common in extragalactic astronomy for correlations between physical quantities to exhibit lots of scatter.
In the BTFR we seem to have a relation with little or no intrinsic scatter.  There are some possible couple of caveats
to this.  First, it is conceivable that the uncertainties in the data have been overestimated.  
Second, selection effects might limit the sample to only those objects that obey the BTFR.

Overestimation of the uncertainties seems unlikely.  That $\chi_{\nu}^2 \approx 1$ is a sign not only that
the model is a reasonable description of the data, but also that the uncertainties have been neither over
nor underestimated.  Even if the data were arbitrarily accurate, so that all of the observed scatter is intrinsic,
the intrinsic width of the relation is still small: the observed 0.24 dex.  If anything, this would appear to
confirm that the disks of the selected galaxies are well behaved dynamically cold rotators.  
If it were otherwise, e.g., if there were substantial non-circular motions in the disks of these galaxies,
then this would be a substantial source of intrinsic scatter and the observed scatter would be much larger \citep{FdZ,trach}.

It is more difficult to address the issue of potential selection effects.  However, in the time since \citet{zwaanTF},
the persistent surprise has been how tight the Tully-Fisher relation remains when challenged with data for new types of galaxies
that we would naively expect to deviate from it.  Since this exercise has now been performed by many different observers
for many different types of galaxies, it seems unlikely that selection effects play a significant role in suppressing the 
scatter of the BTFR.
We select gas rich galaxies here, but it is already well established that star dominated galaxies obey the BTFR
with little scatter \citep[e.g.,][]{verhTF,M05,noordTF}.  The only further selection made here is on data quality.  
When lower quality data are admitted, the scatter increases (Fig.~\ref{btffancyseb} and \ref{scatter}), just as one would expect.  
There is no clear evidence for intrinsic scatter beyond that induced by observational uncertainty and the systematic
effects inevitable in astronomical data.

It is impossible to estimate the scatter that might result from including objects that are not already included.
We have restricted our discussion to rotating galaxies, since these are the objects that define the Tully-Fisher relation.
It is not obvious that it should apply to non-rotating galaxies, though the Faber-Jackson relation
for elliptical galaxies provides an obvious analog.  Dwarf spheroidal galaxies fall remarkably close to the BTFR
with a simple estimate of the equivalent rotation velocity \citep[$V_f = \sqrt{3}\sigma$:][]{boom} though the ultrafaint
dwarfs fall systematically below it \citep{MWolf}.  Even objects that have no apparent business adhering to to the 
BTFR either do so, or come surprisingly close to doing so (e.g., the HI ring in Leo). 
Among rotating disk galaxies one would expect differences in the baryonic mass distribution to have some impact
on the BTFR, just because $V_f^2 \propto GM/R$.  And yet
disk dominated spirals, bulge dominated early types, diffuse low surface brightness galaxies, and dwarf irregulars
all adhere to the same BTFR in spite of their palpably different mass distributions
\citep{zwaanTF,sprayTF,TVbimodal,MdB98a,CR,verhTF,noordTF}.  Rather than finding greater
scatter as we expand our samples beyond the Sc I galaxies traditionally utilized in distance scale studies, 
we find that rotating galaxies of all types adhere to the same BTFR.  It therefore seems unlikely that there remains
some large, unknown population of galaxies in rotational equilibrium that would induce substantial scatter in it.

\subsubsection{Skew and Kurtosis}

A persistent problem with interpreting astronomical data is whether gaussian statistics actually apply.
One reason for this is that systematic errors often dominate random ones.  We can check whether this
might be an issue here by examining higher order statistics like the skew (lopsidedness) 
and kurtosis (pointiness) of the distribution in the deviations $\delta$ perpendicular to the BTFR.  
If the data are well behaved (i.e., distributed normally), the skew and kurtosis should be small.

For all three combinations of samples in Table \ref{fittab}, both the skew and the
kurtosis are small: $|\alpha_3|$ and $|\hat \alpha_4| < 0.15$.  These samples 
are consistent with normal distributions as they should be if random errors dominate over 
systematic effects.  This is especially true for the combination of the
\citet{stark} and \citet{trach} samples, which have $|\alpha_3| = \hat \alpha_4 = 0.01$.  

For comparison, the sample of \citet{gurov} has $\alpha_3 = 0.63$ and $\hat \alpha_4 = -0.51$ around their best fit.
If we trim the three outlying galaxies discussed in \S\ref{datacomp}, the skew is reduced ($\alpha_3 = -0.15$) but
the kurtosis grows ($\hat \alpha_4 = -1.28$).  The negative kurtosis means that the distribution is flat-topped:  it does
not have a well-defined central peak as it should if the data simply scatter around the correct relation.
These data are not distributed normally.  
This undermines the basic assumption of the least square fitting method, so the 
apparently straightforward fit of \citet{gurov} yields a misleading constraint on the slope.

\section{Discussion}

Understanding the physical basis of the BTFR is central to one of the most fundamental issues
presently confronting extragalactic astronomy.  The BTFR provides a
simple test that in principle can distinguish between \LCDM\ and MOND \citep{milgrom83}.
Here we discuss the interpretation of the BTRF in terms of each paradigm in turn.

\subsection{\LCDM}  
\label{LCDMsection}

The interpretation of the BTFR in \LCDM\ depends on how we relate dark and baryonic mass.
Different assumptions can plausibly be made, leading to rather different interpretations.
Here we start from general considerations, later proceeding to specific models.

\begin{figure*}
\epsscale{1.0}
\plotone{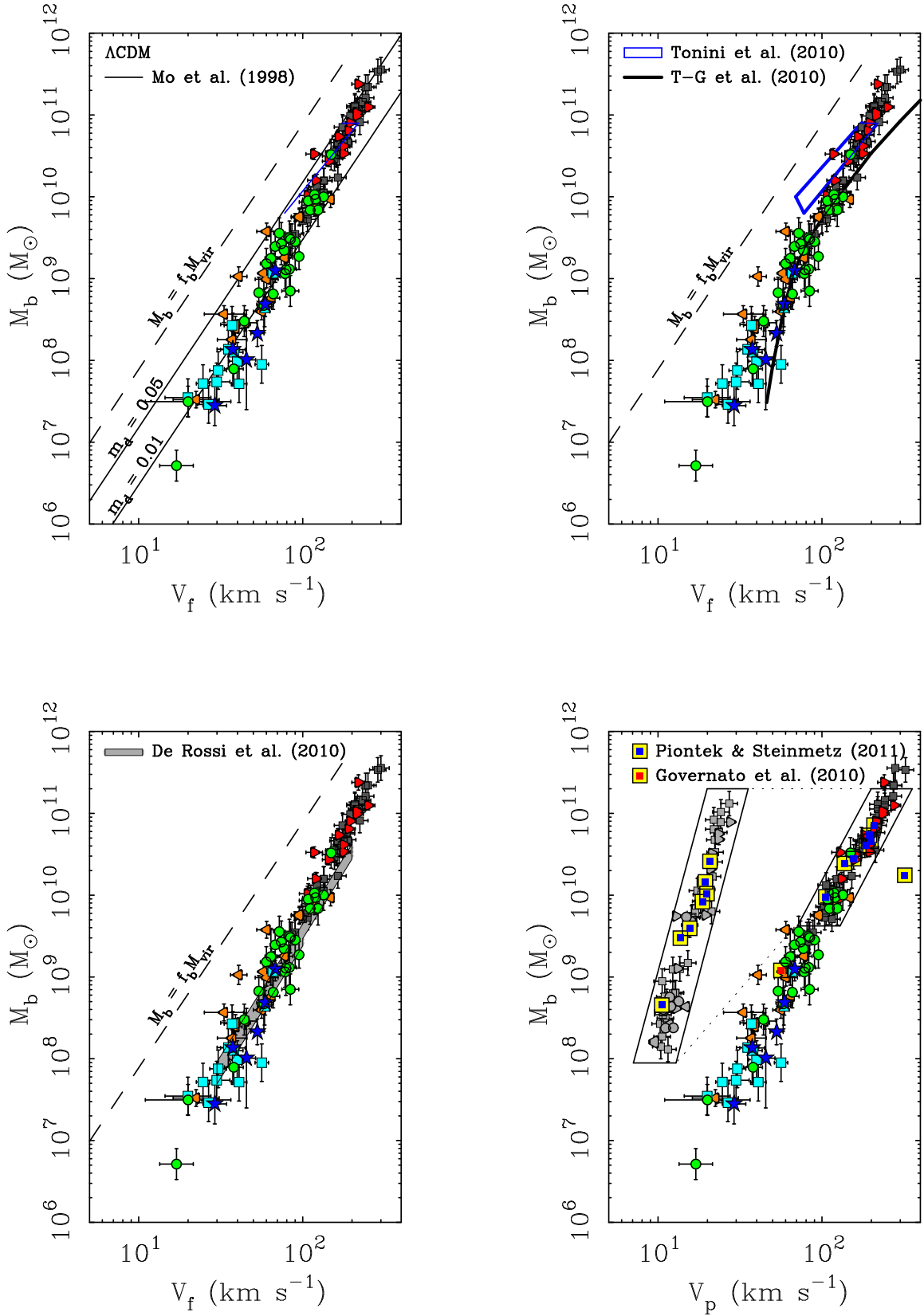}
\caption{The BTFR in \LCDM.  Symbols as per Fig.~\ref{btffancyseb}.
The dashed line illustrates the baryons that are available within the virial radii of dark matter halos with $V_{vir} = V_f$.
Solid lines in the top left panel illustrate the model of \citet{MMW98} for two choices of their disk fraction parameter $m_d$.
In the top right panel, the curved solid line is the model of \citet{TGKPR} while the box illustrates the range of the model
of \citet{TMZBTDS}.  The bottom left panel shows the model of \citet{DRTP}.  The bottom right panel plots the model galaxies
of \citet{piontek} and \citet{fabio} along with the data.  Here the peak rotation velocity $V_p$ is used, as this is more directly 
comparable to the models of \citet{piontek}.  The inset shows that these agree quite well with the data for star dominated
galaxies.
\label{btfcdm}}
\end{figure*}

\placefigure{btfcdm}

It is conventional in cosmology to refer to structures by the density contrast
they represent with respect to the critical density of the universe.  The mass
enclosed within a radius encompassing the over-density $\Delta$ is
\begin{equation}
M_{\Delta} = \frac{4 \pi \Delta}{3} \rho_{crit} R_{\Delta}^3.
\end{equation}  
With $\rho_{crit} = 3H_0^2/8\pi G$, this becomes 
\begin{equation}
M_{\Delta} = \frac{\Delta}{2G} H_0^2 R_{\Delta}^3.
\end{equation}  
By the same token, the circular velocity of a tracer particle at $R_{\Delta}$ is 
$V_{\Delta}^2 = {GM_{\Delta}}/{R_{\Delta}}$.
Consequently, 
\begin{equation}
M_{\Delta} = (\Delta/2)^{-1/2} (G H_0)^{-1} \; V_{\Delta}^3.
\end{equation}  
In \LCDM, the density contrast $\Delta \approx 100$ marks the virial extent
of a dark matter halo \citep{ENS}.
For $H_0 = 72\;\textrm{km}\,\textrm{s}^{-1}\,\textrm{Mpc}^{-1}$,
\begin{equation}
M_{vir} = (4.6 \times 10^5 \;\textrm{M}_{\odot}\,\textrm{km}^{-3}\,\textrm{s}^{3}) V_{vir}^3.
\label{darkTF}
\end{equation}  
This includes all mass, dark and baryonic, that reside within the radius $R_{vir}$.

Dark matter halos thus obey a scaling relation reminiscent of the Tully-Fisher relation \citep[e.g.,][]{SN}.  
This is not the same as the observed Tully-Fisher relation, as both mass and velocity refer to a 
radius $R_{vir}$ that is much larger than what is observed.  
In order to map these theoretical quantities into the observed baryonic
mass $M_b$ and circular velocity $V_f$, we need to introduce some factors:
\begin{equation}
M_b = f_d f_b M_{vir}
\label{fdeqn}
\end{equation}
and
\begin{equation}
V_f = f_V V_{vir}.
\label{fVeqn}
\end{equation}
The total mass of baryons available within $R_{vir}$ is $f_b M_{vir}$, where $f_b$ is the 
cosmic baryon fraction \citep[$f_b = 0.17$:][]{WMAP5}.  The factor $f_d$
represents the fraction of these available baryons that contribute to the observed baryonic
mass.  Similarly, the observed velocity need not be identical to $V_{vir}$; hence the factor $f_V$.

\subsubsection{General Constraints}

The BTFR would follow from equation~(\ref{darkTF}) with a slope of $x=3$ and a 
normalization $\log A = 4.74$ if $f_d = f_V = 1$.  This is shown as the dashed line in Fig.~\ref{btfcdm}.
Had this matched the data, it would provide a very satisfactory interpretation \citep{SN}.
Instead, it exceeds the data, leading us to consider models with $f_d < 1$ and $f_V \ne 1$.  

Equations~\ref{darkTF}-\ref{fVeqn} to provide a generic constraint on \LCDM\ galaxy formation models.
These typically invoke feedback to suppress the cold baryon content of disks so that $f_d < 1$.
We define a feedback efficacy parameter
\begin{equation}
\log\mathcal{E} \equiv 3 \log f_V - \log f_d.
\label{eqn:efficacy}
\end{equation}
When $f_V = 1$, $\mathcal{E}$ is just the inverse of $f_d$, so a galaxy with $\mathcal{E} = 10$ has lost 90\%
of its baryons, and one with $\mathcal{E} = 100$ has lost 99\% of them.  By ``lost'' we mean that they are not part
of the disk of stars and cold gas that contribute to $M_b$ in the BTFR.  It is irrelevant whether these baryons have
been physically expelled as envisioned in many feedback models, or remain in the halo in some dark, unobserved form.
$\mathcal{E}$ merely tells us how effective feedback (or any other mechanism) has been in suppressing the
inclusion of the available baryons into the observed disk.

In addition to variations in$f_d$, the mapping of the observed to virial velocity $f_V$ can also play a role.
This ratio can easily vary with mass as more massive disks boost $f_V$ by engendering more adiabatic
contraction in their host dark matter halos than low mass disks \citep[e.g.,][]{BullockTF}.  My own experience
with modeling adiabatic contraction \citep{adiabat} leads me to expect this to be a modest effect, most likely
in the range $1 \le f_V \le 1.3$.  Very recently, \citet{reyeslens} have combined dynamical and gravitational
lensing measurements to estimate $f_V \approx 1.3$ for galaxies with $M_* > 6 \times 10^9\;M_{\sun}$.  
This represents the upper end of the BTFR (roughly $V_f > 100\;\kms$).  In order to help with the observed variation in
$\mathcal{E}$, $f_V$ would need to be substantially less in lower mass galaxies.  While these often have
rising rotation curves (so possibly $f_V < 1$), flatness of the rotation curve is an explicit selection criterion
of the \citet{stark} sample.  For these galaxies, $f_V \approx 1$ seems the best guess, as their disks are
not massive enough to substantially alter their dark matter halos to yield $f_V > 1$, nor does it seem likely
that the rotation curve will resume rising after being observed to flatten out so as to yield $f_V < 1$.
For the very low mass galaxies of \citet{begum}, $f_V < 1$ is certainly possible, but it seems unlikely to be tiny.
If $f_V = 0.9$ at $V_f = 30\;\kms$ and 1.3 at $300\;\kms$, the total variation in $\mathcal{E}$ is a factor of 3.
This does not, by itself, suffice to explain the data.

Equation~\ref{eqn:efficacy} maps between the parameters of the dark matter halo and those of the
disk galaxy it hosts.   It is completely general, encompassing what needs to happen
to connect theory with observational reality.  In order to be consistent with the data presented here,
models must meet the requirement
\begin{equation}
\log\mathcal{E} = 1.2 - \log \left(\frac{V_f}{100\;\mathrm{km}\,\mathrm{s}^{-1}}\right) -\frac{1}{2} \log \left(\frac{\Delta}{100}\right).
\label{finetuning}
\end{equation}
This constraint applies to models of disk galaxies over the range $30 < V_f < 300\;\kms$.
In applying this constraint, care should be taken to use consistent definitions of total mass
($M_{\Delta}$ with $\Delta = 100$) and  rotation velocity.  The term involving $\Delta$ is a simple
shift in normalization that translates to other common definitions of halo mass (e.g., $\Delta = 200$ instead of 100).  
The constraint holds irrespective of the particular choice of definition for halo mass.

Equation~\ref{finetuning} is a scaling between $\mathcal{E}$ and rotation velocity.
That the efficacy of feedback scales with velocity is not surprising, as the shallower
potential wells of low mass halos should be less effective at retaining baryons.
The efficacy of feedback should also depend on the physics driving it.
For low mass systems like those under consideration here, feedback is often imagined
to be driven by star formation and subsequent supernova events.
In this case, we might expect some correlation between $\mathcal{E}$ and measures of star formation.

\begin{figure*}
\epsscale{1.0}
\plotone{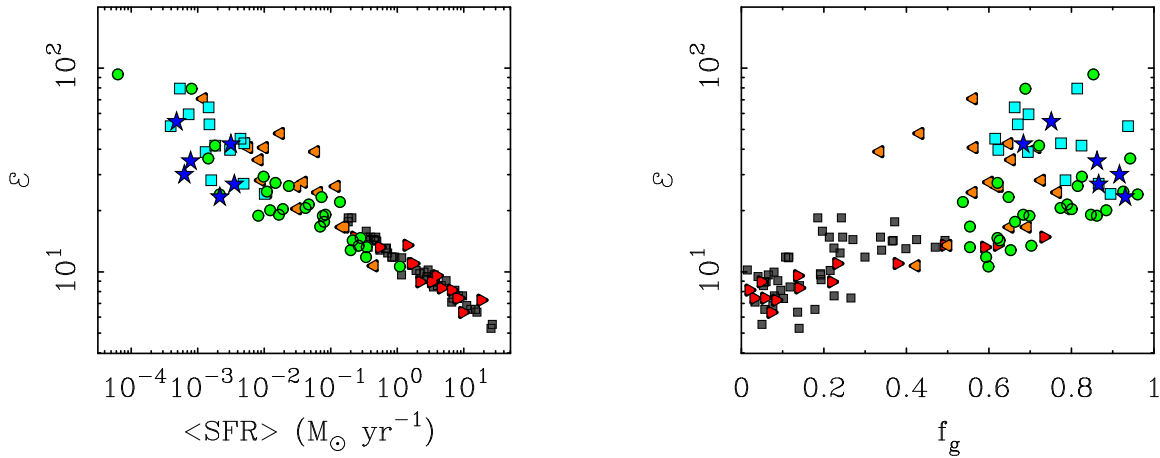}
\caption{The efficacy of feedback (equation~\ref{eqn:efficacy}) as a function of past average star formation rate 
(left) and gas fraction (right).  Symbols as per Fig.~\ref{btffancyseb}.
Feedback solutions reconciling \LCDM\ with the BTFR require galaxies that have 
experienced the least star formation to have been most effective in driving out baryons.  This holds even in galaxies
where there are almost no stars ($f_g \rightarrow 1$).
\label{efficacy}}
\end{figure*}

Fig.~\ref{efficacy} shows $\mathcal{E}$ as a function of past average star formation rate and gas mass
fraction.  The gas fraction $f_g = M_g/M_b$ while to obtain the past average star formation rate we
assume all galaxies have an age of 12 Gyr so that $\langle\mathrm{SFR}\rangle = M_*$/(12 Gyr).
Both of these quantities correlate with $\mathcal{E}$, albeit with substantial scatter at low 
$\langle\mathrm{SFR}\rangle$ and high $f_g$.  The curious thing about these correlations is
that feedback has been most effective in removing or suppressing baryons in galaxies that have
experienced the least star formation and have the most cold gas remaining.  This would seem to
be the opposite of what one would expect from a picture in which intense star formation results
in winds that drive out baryons.  

The effect is not subtle.  Galaxies with $f_g \approx 0.9$ have
$\mathcal{E} \approx 25$; those with $\langle\mathrm{SFR}\rangle \approx 10^{-3} \; \mathrm{M}_{\sun} \, \mathrm{yr}^{-1}$
have $\mathcal{E} \approx 40$.  It appears that the vast majority of baryons associated with the halos of these galaxies
have been ejected or otherwise suppressed.

A more remarkable thing is that $\mathcal{E}$ scales so precisely with $V_f$.  
In order not to impart scatter to the BTFR, there must be very little scatter in equation~\ref{finetuning}.  
I would expect feedback from supernovae to be a chaotic process:
the lack of scatter in $\mathcal{E}$ seems unnatural.  A halo of a given circular velocity could easily host galaxies
of very different $\mathcal{E}$.  Apparently, they do not --- all halos of the same $V_f$ have basically the same $\mathcal{E}$.

The tight relation between $V_f$ and $\mathcal{E}$ provides strong constraints on models of galaxy formation.
Indeed, it seems to me that it constitutes a serious fine-tuning problem.  One can readily envision
models that obtain equation~\ref{finetuning} as a mean scaling relation \citep[e.g.,][]{stringer}.  However, doing so with
so little scatter is a much tougher requirement.  I would expect $f_d$ and $f_V$ to fill out some 
distribution of values \citep[e.g.,][]{kauffmann99,BullockTF}, even for halos of fixed mass.  
There is no obvious reason why these distributions should be either narrow or closely coupled.  
Yet they must be in order to avoid introducing substantial scatter into the BTFR.

\placefigure{efficacy}

Stochastic feedback is not the only expected source of scatter in the BTFR.   
The scatter in the concentration--circular velocity relation for dark matter halos is another.
Without some tuning \citep{BullockTF}, this alone exceeds the scatter allowed by the error budget.

Another expected source of scatter is the triaxiality of dark matter halos \citep[e.g.,][]{KAN}.  
If the gravitational potential is non-symmetric, then the observed velocity depends
on the viewing angle with respect to the principle axis of the potential.   
Otherwise identical galaxies will appear to have difference $V_f$ depending on how
they happen to be oriented on the sky.
A significant degree of non-axisymmetry is clearly excluded \citep{mrandersen,trach,KdN09}.
This constraint might be lessened by the rounding of dark matter halos by the formation of
the baryonic disk, but the disk needs to be heavy enough to have an impact.  \citet{KAN}
find that disks need to be at least half maximal to have a significant impact on their halos.
The low surface brightness disks that make up the majority of the gas rich galaxy sample 
typically fall a factor of two below this level \cite{M05}.

There are other factors though could, and probably should, contribute to the scatter \citep{EL96,MdB98a}.
It is difficult to accommodate all the expected sources of scatter in the error budget. 
This holds even if we have \textit{overestimated} the observational uncertainties, simply because the
observed scatter is small.

The small intrinsic scatter in the BTFR appears to be a serious challenge for \LCDM.
This is hardly the only problem faced by the theory \citep[e.g.,][]{kroupaLG,KdNS11,mia}, so the persistent failure to
detect WIMPs is worrisome.  The \citet{xenon100} has excluded much of the mass--cross
section parameter space where WIMPs were expected to reside \citep{trotta}.
Given the persistence of dynamical puzzles \citep[e.g.,][]{MdB98a,kosowskysellwood},
one might worry that non-baryonic dark matter does not exist after all.

\subsubsection{Some Specific Models}

The generic constraints discussed in the previous section may or may not be adequately
addressed by specific models.  There are very many relevant models in the literature.
It is well beyond the scope of this paper to review all of them.  Nevertheless, it is useful
to examine several illustrative cases.

\citet{MMW98} built a model following the same logic outlined above.  They define a parameter
$m_d$ that is the fraction of the total mass in the disk.  This is analogous to $f_d$, such that $m_d = f_b$ 
when $f_d = 1$.  The resulting Tully-Fisher relation (their equation 4) is shown as lines in the first panel of 
Fig.~\ref{btfcdm} for $m_d = 0.05$ (as they suggest) and for $m_d = 0.01$ \citep[as suggested by][]{gurov}.
The higher value gives a tolerable approximation to high mass spirals while the lower value is appropriate
to gas dominated dwarfs.  A single universal value, as often assumed in galaxy formation modeling, does
not work simply because the slope of the BTFR is steeper than the \LCDM\ halo mass---rotation velocity relation.

More recent models include semi-analytic treatments \citep[e.g.,][]{TMZBTDS,TGKPR} and fully
numerical results \citep[e.g.,][]{MayerMoore,fabioTF,DRTP,fabio,piontek}.
These are shown in the remaining panels of Fig.~\ref{btfcdm}.  All invoke feedback in some way, 
but the prescription for feedback is specific to each so some come closer to matching the constraint
on $\mathcal{E}$ than others.

The model of \citet{TMZBTDS} exhibits a substantial scatter and dependence on the distribution
of stellar mass.  This is natural: the distribution of baryonic mass should matter \citep{MdB98a}.
This model matches the high mass end of the observed BTFR reasonably well, but
over-predicts the mass in galaxies with $V_f \approx 100\;\kms$.
Nevertheless, this model, like that of \citet{MMW98}, provides an honest example of what we
most naturally expect from \LCDM: a substantial amount of scatter about a relation with a slope
of approximately 3.

The model of \citet{TMZBTDS} also indicates a shift in normalization of the relation for galaxies
residing in sub-halos relative to that of primary galaxies (their Fig.~6).  This shift is already apparent at 
relatively large $V_f \gtrsim 100\;\kms$.  There is no
evidence of this in the data, which includes both primary galaxies and dwarfs that are sometimes
isolated and sometimes group members (and hence sub-halos).  For example, KK98-251 is a
dwarf Irregular satellite of NGC 6946, but both lie on the same BTFR with no indication of a shift
in normalization.  Current data for the ultrafaint dwarf satellites of the Milky Way do place them
below the BTFR, but this shift occurs at much smaller scales \citep[$V_f \lesssim 20\;\kms$][]{walker,boom,MWolf}.  

The model of \citet{TGKPR} exhibits significant curvature.  It matches the data well at 
$V_f \sim 100\;\kms$ but falls below it at both high and low $V_f$.  There is no evidence
in the data for curvature in the BTFR of rotating galaxies.
It is of course conceivable that curvature may appear beyond the bounds of the current data,
and may already be observed if one considers dwarf Spheroidals along with dwarf Irregulars
\citep{MWolf}.  This occurs at a significantly lower velocity scale ($\sim 20\;\kms$) 
than in the model of \citet[$\sim 70\;\kms$]{TGKPR}.

The model of \citet{DRTP} fares better, being in rather good agreement with the bulk of the data for
gas rich galaxies.  Like most \LCDM\ models, its slope is relatively shallow ($\sim 3.2$; their Table 3),
but this is only problematic at the high mass end where the model under-predicts the mass of
star dominated galaxies.  This might be remedied by adopting a different IMF, but the change 
would have to be substantial: a factor of 3 or so in stellar mass.  The scatter is respectably small.

Most recently, detailed numerical simulations have produced model galaxies consistent with the BTFR
\citep{fabio,piontek}.  These are shown in the final panel of Fig.~\ref{btfcdm}.  Here we plot the velocity
measured at the peak of the baryonic rotation curve $V_p$ rather than $V_f$ because this is closer to
what is reported by \citet{piontek}.  Even though there is not a great deal of difference between these two
measures of rotation velocity, the already good agreement between the models and the data shown by
\citet{piontek} improves even further when $V_p$ is utilized.  

Numerical simulations of this sort remain computationally expensive.
\citet{piontek} present eight model galaxies, which at present is a large number for such models \citet{fabioTF}.
Of these eight, seven are very nicely consistent with the observed BTFR.  As they discuss, the other case is very 
discrepant.  Obviously, no conclusions can be drawn from one case in eight.  But it does raise some
obvious and important questions.  What is the intrinsic scatter of the BTFR?  
How often we should expect to see grossly discrepant cases?  Would we recognize these objects as normal
disk galaxies?

Resolution also remains an issue in numerical modeling.  The model galaxies of \citet{piontek} are all in the
star dominated regime.  Only the one model galaxy of \citet{fabio} resides in the regime of gas dominated
galaxies where the comparison is most clean.  Comparison with the Millenium simulation suggests that 
simulated galaxies might well follow the observed slope and normalization of the BTFR down to low rotation 
velocities (White \& Guo 2011, private communication).  The critical issue of the intrinsic scatter that should be
exhibited by low mass model galaxies in such simulations remains open.

Detailed numerical simulations appear to have finally reached the point where they can produce quantitatively
realistic model galaxies.  This is clearly a great success.  It remains to be seen if they can provide a satisfactory 
explanation for the small intrinsic scatter in the BTFR.  If they can, then the question becomes whether feedback
prescriptions that accomplish this feat correspond to how feedback works in nature.

\subsection{MOND}

Among alternatives to dark matter, one, MOND \citep{milgrom83}, makes strong \textit{a priori}
predictions about the BTFR, which is a consequence of the form of the force law in MOND.  
This deviates from purely Newtonian at small accelerations, $a \lesssim a_0$, where $a_0$ is
the one new parameter introduced by the theory.  The value of $a_0$ must be constrained by observation
\citep[$a_0 \approx 10^{-10}\;\textrm{m}\,\textrm{s}^{-2}$:][]{SMmond} 
but once specified is constant.  

In the deep modified regime, $a \ll a_0$,
the effective acceleration $a \rightarrow \sqrt{g_N a_0}$, where $g_N$ is the Newtonian
acceleration calculated for the observed baryonic mass in the usual way.  For circular motion around a
point source, we can equate the centripetal acceleration to this effective force to obtain
\begin{equation}
a_0 G M = V_f^4.
\label{eqn:mondTF}
\end{equation}
Since both $a_0$ and $G$ are constants, and all mass is baryonic,
one recognizes the BTFR, $M_b \propto V_f^4$.

There are several consequences of equation~\ref{eqn:mondTF} \citep{milgrom83}.
The Tully-Fisher relation is absolute, being a direct consequence of the force law.
There should be no intrinsic scatter.  The normalization is set by constants of nature.
Contrary to the case in Newtonian dynamics, the circular velocity does not depend on the 
radial extent of the mass distribution, so there is no dependence on surface brightness or scale length. 
This naturally explains the lack of observed size or surface brightness
residuals from the Tully-Fisher relation \citep{zwaanTF,sprayTF,CR,myPRL}.

Indeed, \citet{milgrom83} specifically predicted that low surface brightness galaxies would share the same
Tully-Fisher relation, with the same normalization, as high surface brightness spirals.  This prediction
was subsequently confirmed \citep{MdB98b}.  The data for gas rich galaxies now provides the opportunity 
to test another \textit{a priori} prediction.  

\subsubsection{A Test with No Free Parameters}

The circular velocities of the gas rich dwarfs considered here are measured. 
Their baryonic masses are observed.  Their location on the BTFR is fixed.
Consequently, the gas rich dwarfs provide a test of the MOND
prediction with \textit{zero} free parameters.  

The results of this test is shown in Fig.~\ref{btfmond}.
Even though the value of $a_0$ was set by early fits to bright spirals 
\citep[$a_0 = 1.2\;\textrm{\AA}\,\textrm{s}^{-2}$:][]{BBS},
equation~\ref{eqn:mondTF} fits the data for much lower mass galaxies well.
The gas rich dwarfs follow the prediction of MOND with no fitting whatsoever \citep{newPRL}.

There was no guarantee that MOND would pass the test posed by gas rich galaxies.
A Tully-Fisher type relation is built into the MOND force law, but its slope is fixed and its normalization
is constrained by star dominated galaxies.  The data for gas rich galaxies could have fallen anywhere
in the BTFR plane.  They might have exhibited the slope of 3 that is more natural to \LCDM.   
Instead, they fall along the one and only line permitted in MOND.

\subsubsection{The Value of $a_0$}

The data for gas rich galaxies are consistent with existing estimates of the acceleration constant $a_0$
\citep{SMmond}.  We can reverse the question, and ask what the best value of $a_0$ is according to these
data.  The best fit intercept of the BTFR already contains this information, and the corresponding value of
$a_0$ is given in Table \ref{fittab}.

There are some subtle issues to consider in calibrating $a_0$.  This has previously been done by fits
to full rotation curves \citep[e.g.,][]{BBS,swatersmond}.  These contain more information than simply
the outer circular velocity, but require an additional assumption.  In particular, a form for the
function that interpolates between the Newtonian and MOND regimes must be assumed.  For an
assumed form, the best fit mass-to-light ratio must be found simultaneously with the best fit value of
$a_0$.  There is not much range allowed in either of these parameters, but there can be  
covariance between them, so it is challenging to disentangle them uniquely.  Here, the data are not
so precise, and we make no use of the information in the shape of the rotation curve, only its outer
value.  However, this approach has the virtue of making a minimum of 
assumptions and using mass estimates that are completely independent of the theory.

Strictly speaking, equation \ref{eqn:mondTF} applies at infinite distance from an isolated point mass.
Real galaxies are not point masses, and $V_f$ is measured at a finite radius.  Consequently, the
intercept of the empirical BTFR is related to the acceleration constant of MOND by
\begin{equation}
A = \frac{\chi}{a_0 G}
\label{eqn:Aa0}
\end{equation}
where $\chi$ is a factor of order unity that accounts for the finite size and non-spherical geometry
of galaxies.

\citet{MdB98b} estimated $\chi = 0.76$ from purely geometrical considerations --- thin disks rotate faster
than the equivalent spherical mass distribution.  The sample discussed
by \citet{M05} has MOND fits performed for $a_0 = 1.2\;\textrm{\AA}\,\textrm{s}^{-2}$ and a separate fit
for the BTFR yielding $A = 50\;\Aunits$.  This gives an empirical estimate of $\chi = 0.80$ which I adopt
here.  The gas rich galaxy data thus give
\begin{equation}
a_0 = 1.3 \pm 0.3\;\textrm{\AA}\,\textrm{s}^{-2}.
\label{a0value}
\end{equation}
The uncertainty here includes the formal uncertainty in $A$ (equation \ref{bestfitA})
plus an allowance for 20\% uncertainty in $\chi$.  It is conceivable that there could be a small
systematic offset in the mean value of $\chi$ between gas and star dominated galaxies.

\begin{figure}
\epsscale{1.0}
\plotone{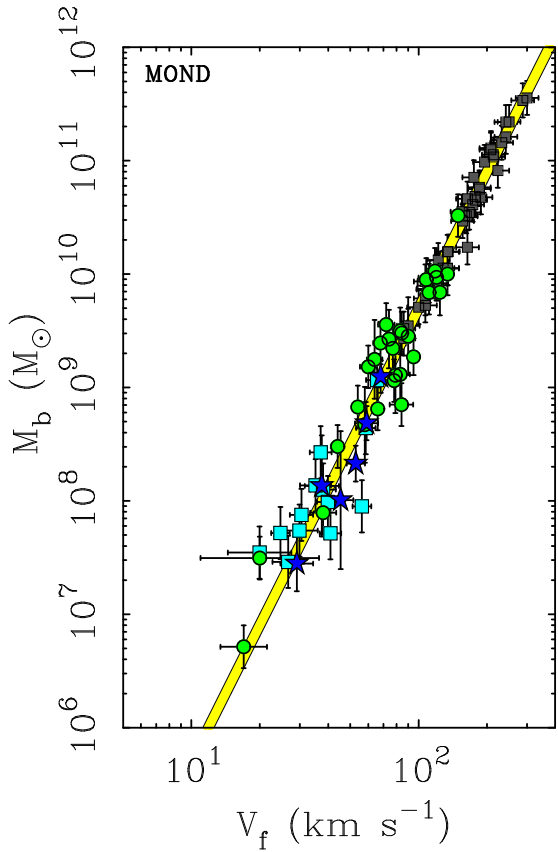}
\caption{The BTFR in MOND.  Symbols as per Fig.~\ref{btffancyseb}.
The band shows the expectation of MOND \citep{milgrom83}. 
The width of the band represents the $\pm 1 \sigma$ uncertainty in $a_0$ based on detailed 
fits to the rotation curves of a subset \citep{BBS} of the star dominated galaxies (gray squares).  
The gas rich galaxies test the MOND prediction in a new regime with zero free parameters.
\label{btfmond}}
\end{figure}

\placefigure{btfmond}

Detailed rotation curve fits to some of the same galaxies suggest 
$a_0 \approx 0.7\;\textrm{\AA}\,\textrm{s}^{-2}$ \citep{swatersmond}.
This becomes $a_0 \approx 1.0\;\textrm{\AA}\,\textrm{s}^{-2}$ if we exercise the same quality
control on the data as applied here.
This gives some idea of the systematic uncertainty and the impact of the best-fit stellar mass-to-light
ratio when it is treated as a fitting parameter along with $a_0$.
However, data quality remains the paramount issue.

The BTFR is clearly a success of MOND, despite its many other drawbacks.
Further tests are possible, as adherence to the BTFR does not guarantee that it is possible to obtain an
acceptable fit to the entire rotation curve of each galaxy.  While MOND is known to do quite well in some
gas rich galaxies \citep[e.g.,][]{N1560}, it may have problems in others \citep[e.g.,][where again
the issue appears to be with the inclination]{sanchez}.  It is particularly challenging to fit the full
rotation curve of NGC 3198 \citep{bottemamond,gentileTHINGS}, 
the most massive galaxy to meet the gas rich criterion imposed here.  
Perhaps this suffices to falsify MOND, but we should not be so eager to disbelieve MOND as a theory
that we ignore the remarkable empirical virtues of the simple formula suggested by \citet{milgrom83}.

There are other suggestions for modifying gravitational theory besides MOND. 
Conformal Weyl gravity \citep{weyl} and MoG \citep{MoG} offer two examples.  
If these theories can fit rotation curves \citep{BrownMoff,weylfits}, then they will also fit the BTFR
\citep{MoGBTF}.  However, I am not aware of a clear \textit{a priori} prediction of this phenomenon 
by any theory other than MOND.

\section{Conclusions}

The baryonic mass--velocity relation provides an important test of \LCDM\ and alternatives like MOND.
I have assembled recent data for gas rich rotating galaxies in order to test these theories.
With $M_g > M_*$, the location of these objects in the baryonic mass--rotation velocity plane is effectively measured
directly from observations without the usual systematic uncertainty in the stellar mass-to-light ratio.

A long standing prediction \citep{milgrom83} of MOND is that rotating galaxies will fall on a single
mass--velocity relation with slope 4:  $M_b \propto V_f^4$.  This prediction is realized in multiple
independent data sets.  Gas rich galaxies fall where predicted by MOND with no free parameters.
There are not many predictions in extragalactic astronomy that fare so well
over a quarter century after their publication.  

In \LCDM, the naive expectation of a of Tully-Fisher type relation with a slope around 3 fails to predict
the location of gas rich galaxies in the plane of the BTFR.  
Only the most recent detailed numerical models \citep[e.g.,][]{fabio,piontek} succeed in reproducing the 
observed phenomenology.  This is certainly progress.  However, a physical understanding for why galaxy
formation in the context of \LCDM\ should pick out the particular phenomenology predicted \textit{a priori} by 
MOND remains wanting.  

\acknowledgements  The work of S.S.M.\ is supported in part by NSF grant AST 0908370.  



\end{document}